\documentclass[12pt,preprint]{aastex}

\def\lsim{\raise0.3ex\hbox{$<$}\kern-0.75em{\lower0.65ex\hbox{$\sim$}}}
\def\gsim{\raise0.3ex\hbox{$>$}\kern-0.75em{\lower0.65ex\hbox{$\sim$}}}

\begin{document}


\title{Low $R_V$ from circumstellar dust around supernovae}


\author{Ariel Goobar}
\affil{Physics Department, Stockholm University,
    AlbaNova University Center, SE 106 192 Stockholm, Sweden}

%




\begin{abstract}
The effective extinction law for supernovae surrounded by
circumstellar dust is examined by Monte-Carlo simulations. Grains with
light scattering properties as for interstellar dust in the Milky-Way
(MW) or the Large Magellanic Clouds (LMC), but surrounding the
explosion site would cause a semi-diffusive propagation of light up to
the edge of the dust shell. Multiple scattering of photons
predominantly attenuates photons with shorter wavelengths, thus
steepening the effective extinction law as compared to the case of
single scattering in the interstellar medium. Our simulations yield
typical values for the total to selective extinction ratio $R_V\sim
1.5-2.5$, as seen in recent studies of Type Ia supernova colors, with
further stiffening differential extinction toward the near-UV. 
\end{abstract}





\section{Introduction}
The uncertainties in the brightness corrections of Type Ia supernovae
(SNIa) for color excess is among the largest systematic uncertainties
in the use of Type Ia supernovae to measure cosmological distances
\citep{smock}. The standard interpretation of color excess being due
to extinction by interstellar dust in the supernova host galaxy has
recently been challenged by the empirically deduced
color-brightness relation for SNIa. While studies of differential
extinction of quasars shining through foreground galaxies
yield values of $R_V= A_V/E(B-V)$ compatible with the 
average MW value \citep{QSO}\footnote{It should
be noted that low $R_V$ values to individual QSO systems have 
been found in e.g. \citep{QSO} and \citep{Wang04}}, the SNIa Hubble diagram scatter is
minimized for values significantly
smaller than  $R_V=3.1$ \citep{SNLS,scp08}. Furthermore, the use
of SNIa for cosmology benefits significantly from the 
understanding the extinction law for the full optical
and near-IR wavelength range. E.g. the wavelength dependence 
of extinction toward the Magellanic Clouds differ from the Milky-Way
extinction law in \citep{CCM}, even for very similar values of $R_V$.  
For quasar sight-lines, \citet{QSO} tested both Milky-Way like
extinction as well as Small Magellanic Cloud (SMC) extinction law, both
giving comparable goodness of fit. A preference for SMC dust for
extinction of AGNs has been suggested by \citet{li}.

Recently, the detection of circumstellar (CS) matter in the local
environment surrounding the Type Ia supernova 
SN2006X in the nearby galaxy M100 has been reported by \citet{Patat07}. A shell within a few
$10^{16}$ cm ($\sim 0.01$ pc) of the center of the explosion has been
suggested to explain the time-variable Na I D lines in the SN spectrum.

\citet{Wang08a} report $R_V=1.48 \pm 0.06$ and $E(B-V) = 1.42 \pm
0.04$ mag for SN2006X and a light echo in the lightcurve was found by
\citet{Wang08b} consistent with dust illuminated at a distance of
27-170 pc from the site of the explosion. Even if the local
environment around this supernova may not be very common among SNIa,
similar values for the total to selective extinction ratio have
been reported for several SNIa with good wavelength coverage. E.g.
\citet{Krisciunas07} found $R_V=1.55 \pm 0.08$ for SN 1999cl;
\citet{Elias-Rosa06,Elias-Rosa08} report $R_V=1.80 \pm 0.19$ and
$R_V=1.59 \pm 0.07$ for SN2003cg and SN 2002cv respectively. Furthermore,
a
statistical study of optical colors of a sample including 80 near-by SNIa,
\citet{Nobili&Goobar} found an average value of $\bar R_V=1.75
\pm 0.27$ for SNIa with $E(B-V)<$0.7, and even lower for a subsample 
of low-reddening SNIa.

Next, we examine the possibility that low values of $R_V$ stem
from the semi-diffusive propagation of photons in the neighborhood
of the site of the supernova explosion.

\section{Quasi-diffusive light propagation around the supernova}
Photon propagation around a medium of scatterers can be described by
a quasi random-walk picture. The reader
is referred to \citep{chandra} for a beautiful introduction to this
subject.
Lets consider a localized distribution of
dust particles within a distance $R_{CS}$ from the explosion site,
negligibly small compared to the distance to the observer, $d$,
i.e. $R_{SN}< R_{CS} \ll d$, where $R_{SN}$ corresponds to the radius
from where the SN radiation emerges. The trajectory of a photon will 
be straight until it hits a dust particle at which point the photon
can either be scattered or absorbed. If the photon is scattered
in a new direction, it follows a straight path until the next
encounter, and so on until $r>R_{CS}$.
The mean free path between interactions, $\lambda_{eff}$, is thus
determined by the number density of scatterers, $n$, and their
effective cross-section for scattering and absorption
of light, $\sigma_{eff}=\sigma_{s}+\sigma_{a}$:
\begin{equation}
\lambda_{eff} = {1 \over n \cdot \sigma_{eff}}
\label{eq:lambda_s}
\end{equation}

For $\lambda_{eff} \gg R_{CS}$, the single scattering approximation is
valid and the light beam reaching the observer is attenuated
as $e^{-\tau}$, with $\tau=R_{CS}/\lambda_{eff}$. This is the case applicable for
extinction by dust in the interstellar medium.

For $\lambda_{eff} \ll R_{CS}$, corresponding to a (local) high number
density of scatterers, the situation is different.  If the absorption
probability is much lower than for scattering, photon propagation is
diffusive and the average properties can described analytically using
the formulas for random-walk \citep{amanda}. The case we are
considering here is for $\lambda_{eff}\sim R_{CS}$, where the
scattering cross-section exceeds the absorption cross-section ,
$\sigma_s>\sigma_a$, in a wavelength dependent manner. In particular,
we examine the cases where the light scattering properties of dust
particles in the CS matter are similar to what has been modeled for
interstellar dust grains the Milky-Way \citep{Draine03} or the LMC
\citep{WD01}. 

Table~\ref{tbl-1} shows the the wavelength dependent albedo factor ($=
\sigma_{s}/(\sigma_{s}+\sigma_{a})$) and the average of the cosine of
the scattering angle for interactions between light and dust
particles. Also tabulated is the absorption cross-section divided by
dust mass. Note that the Milky-Way parameters correspond to a dust
size distribution matching $R_V=3.1$ for dimming of stars in the
Galaxy. \citet{lifan05} considered the impact of circumstellar dust
upon the measured value of $R_V$, but only the extreme case where all
scattered photons may reach the observer. That assumption overlooks an
important aspect of the problem: while the bluer photons scatter more,
they also are more likely to be absorbed. This leads to a steeper
wavelength dependence of the effective extinction law, possibly
explaining the unusual total to selective extinction ratios
found in studies of SNIa.

\section{Monte-Carlo simulation of light propagation around supernova}
\begin{figure}
\plotone{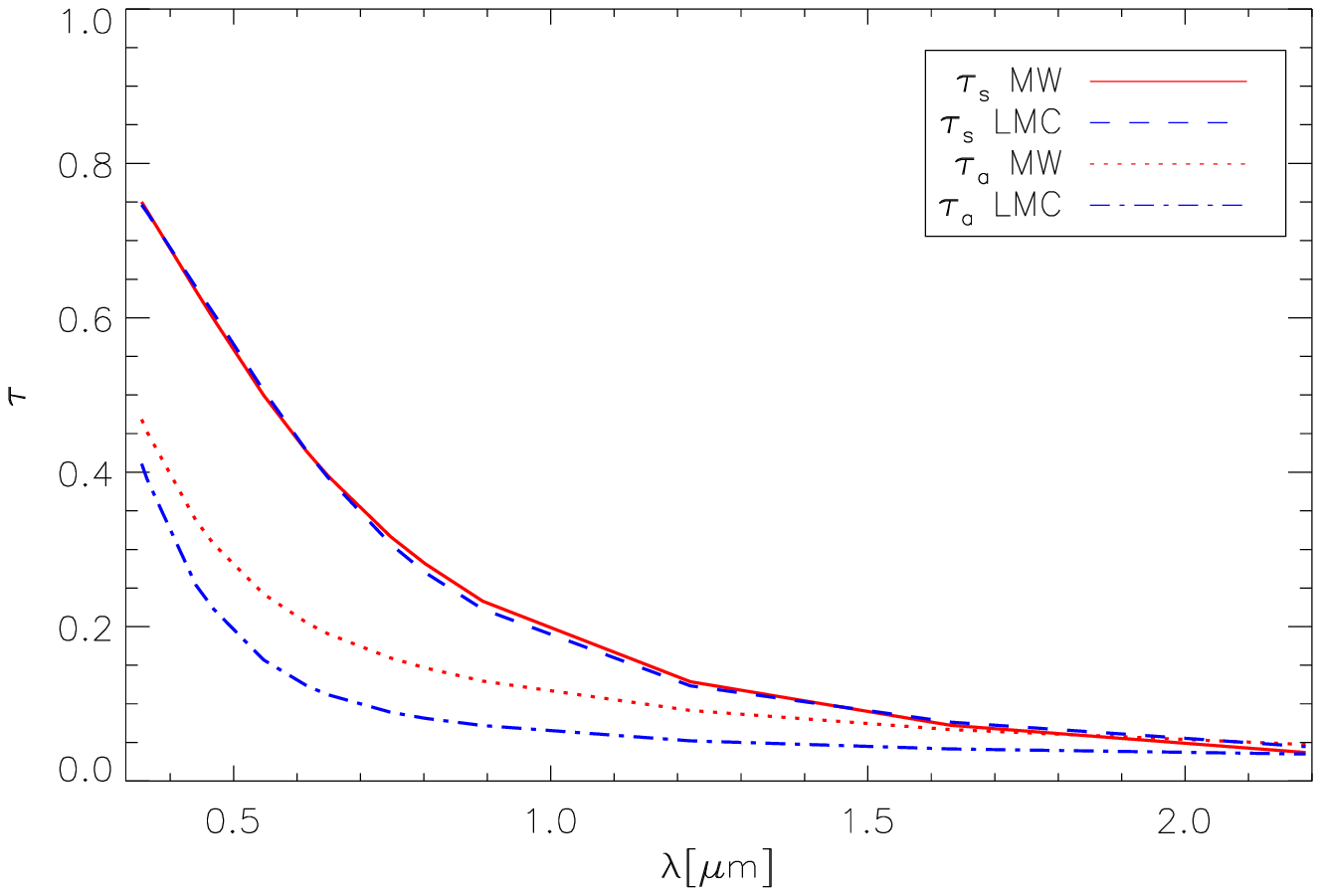}
\caption{Optical depth for scattering ($\tau_s=R_{CS}/\lambda_s$)
and absorption  ($\tau_a=R_{CS}/\lambda_a$) as a function of wavelength 
from the Monte-Carlo simulation of photon propagation 
in a dusty circumstellar material around the SN for $E(B-V)=0.1$
at $r=R_{CS}$. Scattering properties for average MW dust from
\citep{Draine03} and LMC dust \citep{WD01} tabulated in 
Table~\ref{tbl-1} were used. The
bluer photons scatter more and since they also have a larger probability
for absorption, fewer make it to the edge of the dusty shell, $R_{CS}$. 
As a result, a steeper wavelength dependence for attenuation is to 
be expected.
\label{fig:tau}}
\end{figure}

In order to estimate the net effect of scattering and absorption
on the light reaching the outer edge of a shell of circumstellar
dust around the SN site, $R_{CS}$, a Monte-Carlo simulation
was performed. Photons with energies corresponding to
the central wavelengths of the $UBVRIJHK$ photometric system
were generated and subsequently followed as they propagate in the
dusty medium.\footnote{In this study we neglect the effect
noted by \citet{lifan05} where the effective wavelength in each
filter is changing with time following the color evolution of
SNIa}
A uniform distribution
of scatterers within a sphere of radius $R_{CS}$ is used in the
calculations. Our treatment is rather insensitive of the physical size 
of $R_{CS}$ since what governs the differences in path-lengths 
of photons at different
wavelengths are the optical depths, $\tau_s=R_{CS}/\lambda_s$ and
$\tau_a=R_{CS}/\lambda_a$. Thus, for a fixed color excess, a
larger $R_{CS}$ can be compensated by a lower number density, $n$,
thus keeping $\tau_s$ and $\tau_a$ unchanged. 
Figure~\ref{fig:tau} shows 
the wavelength dependence of $\tau_a$ and $\tau_s$ in our
calculation for MW and LMC
dust types for a reddening at $r=R_{CS}$ of $E(B-V)=0.1$. The key observation is
that at short wavelengths, both $\tau_s$  and $\tau_a$
increase, while $\tau_s$ dominates. Since
they scatter more, photons with short wavelength 
leaving the CS region to eventually reach
the observer must propagate a larger path-length than photons 
at longer wavelengths. However, since $\tau_a$ also increases
at shorter wavelengths, the relative attenuation of bluer photons
is enhanced, thus generating an effective extinction law
with steeper wavelength dependence, i.e. lower $R_V$. 

SMC-type dust, as parameterized in \citep{WD01}, has a smaller albedo
in the optical to near-IR region (see Fig.~23 in \citet{WD01}), leading to a
larger value of $\tau_a$ compared to LMC or MW dust. Thus, SMC dust is not 
suitable to explain the anomalous SN colors with the scenario
presented in this work.  

For each photon starting at $R_{SN} \ll R_{CS}$, straight path-lengths
between interactions ($L_s$,$L_a$) were generated from an exponential
distributions $\lambda_s^{-1}\exp{(-L_s/\lambda_s)}$ and 
$\lambda_a^{-1}\exp{(-L_a/\lambda_a)}$, where 
the mean free path for scattering and absorption,
$\lambda_s=(n\cdot\sigma_s)^{-1}$ and
$\lambda_a=(n\cdot\sigma_a)^{-1}$ were calculated from the wavelength
dependent parameters in Table~\ref{tbl-1}.  The scattering angle 
for each interaction was
generated in the simulation following the Henyey-Greenstein
approximation \citep{henyey}:
\begin{equation}
{{\rm d\sigma} \over {\rm d(\cos(\theta))}} = 
{ 
{1 - g^2} 
\over 
{\left(1 + g^2-2g \cos(\theta)\right)^{3 \over 2}}
},
\label{eq:henyey}
\end{equation}  
where $g=<\cos(\theta)>$ (also listed in Table~\ref{tbl-1}).  The
density of dust in the shell was varied to cover a wide range of
reddening at $r=R_{CS}$, $0\le E(B-V)\le 0.4$. The probability for
photons to reach $R_{CS}$ without absorption is calculated by
repeating the ray-tracing Monte-Carlo $5\cdot10^5$ times for each
wavelength.

\section{Low ${R_V}$ from simulations}
\begin{figure}
\plotone{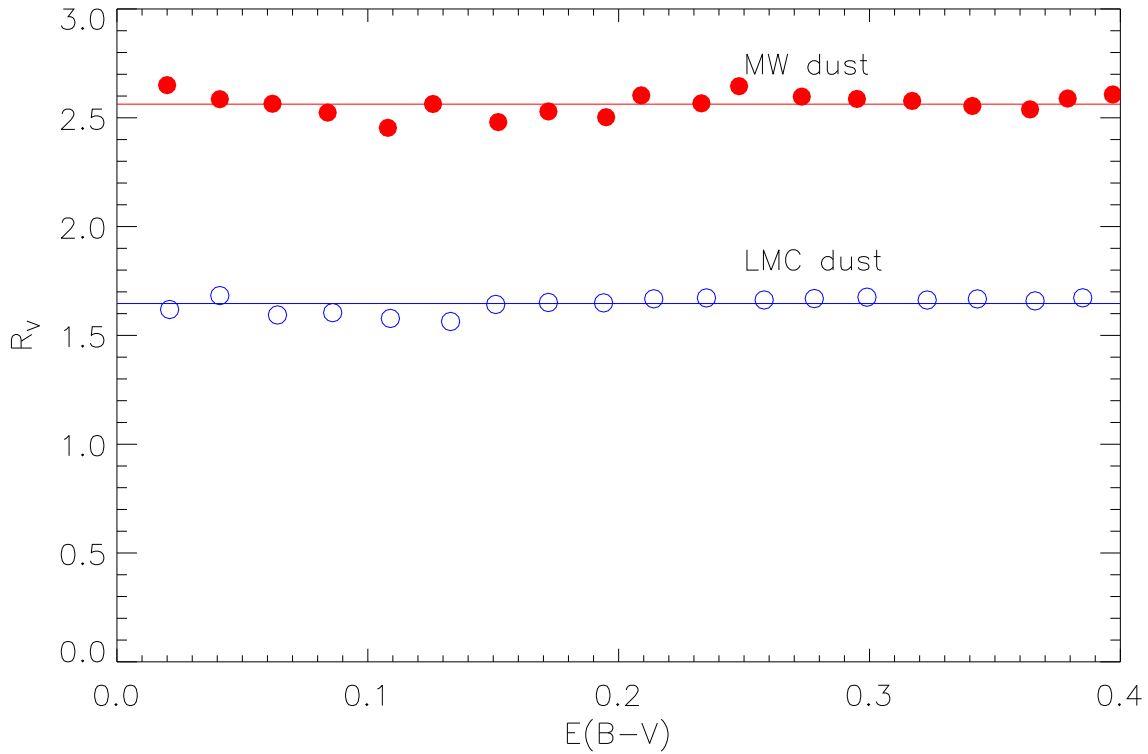}
\caption{$R_V=A_V/E(B-V)$ vs $E(B-V)$ resulting from the Monte-Carlo
simulation of photon propagation in a dusty circumstellar material
around the explosion site. The ``wiggles'' in the curve are compatible
with the statistical uncertainty of the Monte-Carlo
simulation. Scattering properties for average MW dust from
\citep{Draine03} and LMC dust \citep{WD01} tabulated in
Table~\ref{tbl-1} were used. The average total to selective 
extinction ratios found were $R_V=2.56$ for Milky-Way type dust
and $R_V=1.65$ for dust compatible with properties in
the Large Magellanic Clouds.
\label{fig:RVvsEBV}}
\end{figure}

\begin{figure}
\plotone{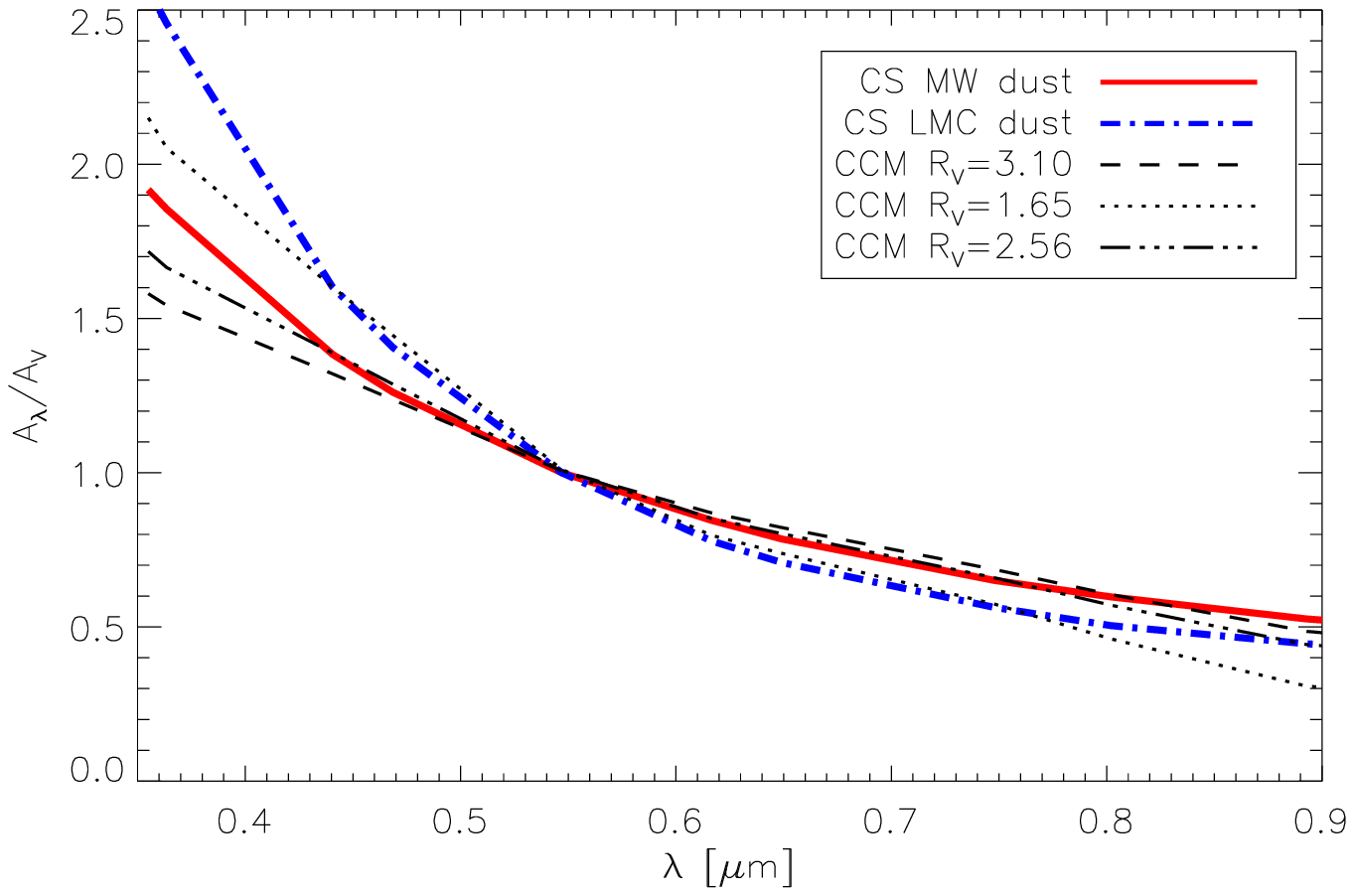}
\caption{The extinction law, $A_\lambda/A_V$, derived from the Monte-Carlo
simulation of circumstellar dust are compared
with the parameterized Milky-Way extinction law \citep{CCM}.
scattering properties for average MW dust from
\citep{Draine03} and LMC dust \citep{WD01} tabulated in
Table~\ref{tbl-1} were used.
The two cases considered yield a poor match to the standard $R_V=3.1$
case. Even when the $R_V$ is adjusted to fit the $A_V/E(B-V)$,
differences are found in the wavelength dependence, especially
at shorter wavelengths.\label{fig:extlaw}}
\end{figure}

Figure~\ref{fig:RVvsEBV} shows that the attenuation of light after the
circumstellar shell yields low values of $R_V=A_V/E(B-V)$, as
anticipated.  In particular, the simulations using LMC dust, result in
an effective $R_V=1.65$, compatible with the best fit result of
\citep{Nobili&Goobar}. We note, however, that the CCM extinction law
\citep{CCM} does not accurately reproduce the wavelength dependence of
photons attenuation over the entire optical range, as shown in
Figure~\ref{fig:extlaw}\footnote{The validity of the simulation code
  was tested by accurately reproducing the standard extinction laws
  when the diffusive process was switched off.}. The deviations are
significant for the U-B color, in agreement with the observations
reported in \citep{salt}.

\begin{figure}
\plotone{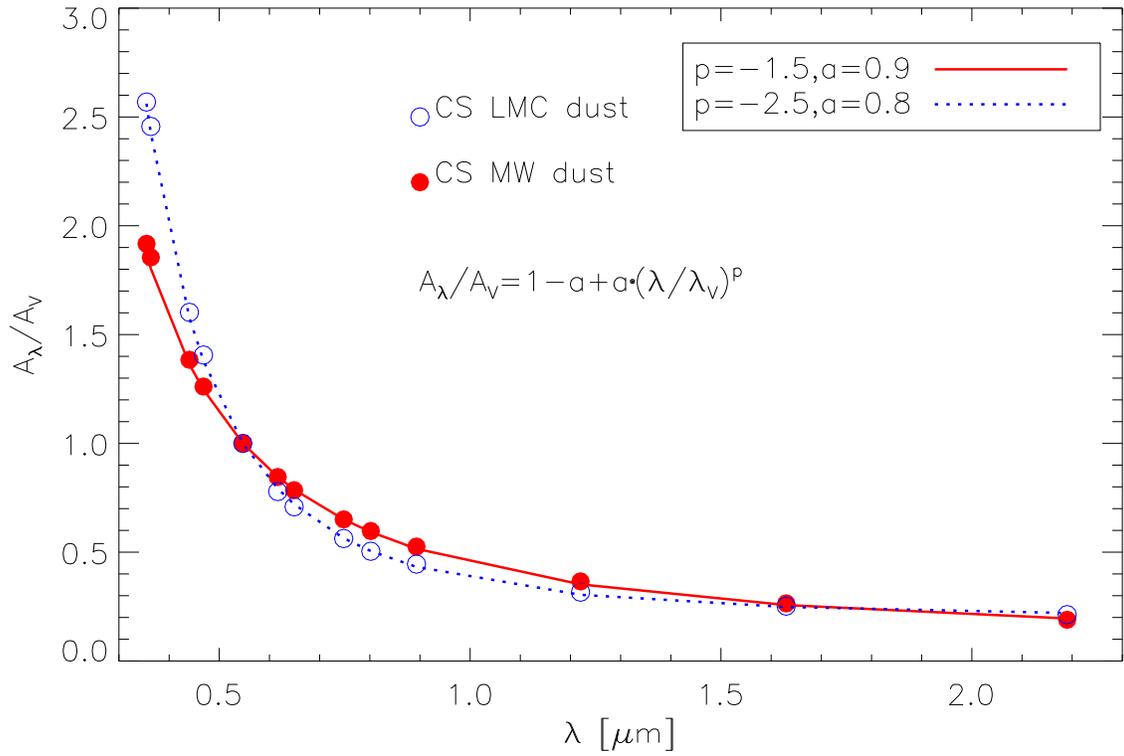}
\caption{Empirical functional fit to the simulated data.
scattering properties for average MW dust from
\citep{Draine03} and LMC dust \citep{WD01} tabulated in
Table~\ref{tbl-1} were used.
\label{fig:fitfun}}
\end{figure}

Instead, a power-law relation gives a good fit to the simulations (see Figure~\ref{fig:fitfun}):
\begin{equation}
{A_\lambda \over A_V} 
= 1 - a + a\left({\lambda \over \lambda_V}\right)^p,
\label{eq:power}
\end{equation}

where $\lambda_V=0.55$ $\mu$m is the central wavelength of the
V-band filter. Since the attenuation at B-band (central 
wavelength $\lambda_B$) is given
by $(R_V+1)E(B-V)$, the parameters in Equation~(\ref{eq:power})
are related to $R_V$ as
\begin{equation}
R_V = 
a^{-1}\left[ \left({\lambda_B \over \lambda_V}\right)^p -1\right]^{-1}\\
 = {1 \over  a \left(0.8^p -1\right)}
\end{equation}
It is encouraging for the use of Type Ia SNe for precision cosmology
that the reddening corrections may take a simple and general
analytical form, if the model is confirmed. Further 
studies of SN colors, also including near-IR data, should be used
to test this relation.

\section{Summary and conclusions}
Simple simulations show that circumstellar material, detected in
at least one Type Ia supernova, SN2006X \citep{Patat07,Wang08b}, could
potentially explain the empirically determined extinction law for low redshift
SNIa, especially if the circumstellar material
resembles LMC dust grains. Adopting the CS shell size of
\citet{Patat07}, $R_{CS}\sim 10^{16}$ cm, we find that for
$\tau\sim1$, the required mass in dust around the supernova is
$M_{dust}\sim 4 \pi R_{CS}^2/(\sigma_a/m_{dust})\sim 10^{-4} M_\odot$
when inserting typical values of the absorption cross-section from
Table~\ref{tbl-1}. 

A simple power-law expression is found to fit very
well the effective extinction law for dust in the CS environment
of the supernova produced by Monte-Carlo simulations.

Depending on the thickness of the CS shell, shifts in
the time of lightcurve maximum may be expected for different bands
since the amount of quasi random-walk will differ. In particular,
photons in redder bands will suffer less scattering and thereby less
time delay. This effect should correlate with the measured reddening,
$E(B-V)$. 
The assumption of a uniform
density is not expected to be critical for the results at first
order. However, a second order effect may be expected since a large
scale of $R_{CS}$ would result in a longer time for photons being
``trapped'' in the scattering sphere. As the the intrinsic colors of
Type Ia change on a time scale of days \citep{Nobili&Goobar}, time
delays of photons of that time scale would affect the measured colors
as a function of time.

In a forthcoming paper, potential direct
observables from interaction between photons and dust will be
investigated. Also, the sensitivity of the effective extinction law to the
dust grain sizes and density profile in CS medium and the combination
of both scattering in the circumstellar material and the interstellar
medium needs to be further investigated.    
If the presence of CS material is indeed the
source of the color-brightness relation found in SNIa, the case for
restframe near-IR observations is further strengthened: the peak
magnitude corrections, and their model dependence, are smaller than at
optical wavelengths.






\acknowledgments

I am grateful to Vallery Stanishev for discussions triggering this work and
to the anonymous referee for providing constructive comments which have
greatly improved the paper. It is a pleasure to thank Christian Walck
for making his random number generation package available to me. Thanks
also to Eric Linder for pointing out a typo in the original manuscript.

\clearpage




\clearpage

\clearpage

\begin{deluxetable}{ccccc}
\tabletypesize{\scriptsize}
\tablecaption{Scattering parameters for optical and near-IR
photons from \citep{WD01,Draine03} corresponding to interstellar
extinction in the Large Magellanic Clouds (LMC) and 
Milky-Way (MW) with $R_V=3.1$. \label{tbl-1}}
\tablewidth{0pt}
\tablehead{
\colhead{wavelength}                 & 
\colhead{albedo}                     & 
\colhead{g=$<{\cos(\theta)}>$}       & 
\colhead{${\sigma_a \over m_{dust}}$(cm$^2/$g)} & 
\colhead{filter} \\
\colhead{($\mu$m)}           & 
\colhead{MW \ \  LMC}                 & 
\colhead{MW \ \  LMC}                  & 
\colhead{MW \ \  LMC}                  & 
\colhead{}
}
\startdata
0.36 & 0.6203 \ \  0.6532 & 0.5695 \ \ 0.6072 & 1.595E+04 \ \  1.162E+04 & U \\
0.44 & 0.6529 \ \  0.7159 & 0.5654 \ \ 0.6153 & 1.191E+04 \ \ 7.542E+03 & B \\
0.55 & 0.6735 \ \  0.7631 & 0.5382 \ \ 0.6059 & 8.551E+03 \ \ 4.666E+03 & V \\
0.65 & 0.6745 \ \  0.7785 & 0.4995 \ \ 0.5815 & 6.722E+03 \ \ 3.319E+03 & R \\
0.80 & 0.6576 \ \  0.7686 & 0.4381 \ \ 0.5289 & 5.170E+03 \ \ 2.416E+03 & I \\
1.22 & 0.5846 \ \  0.7029 & 0.2893 \ \ 0.3892 & 3.224E+03 \ \ 1.550E+03 & J \\
1.63 & 0.5200 \ \  0.6472 & 0.2086 \ \ 0.2911 & 2.351E+03 \ \ 1.237E+03 & H \\
2.19 & 0.4391 \ \  0.5599 & 0.1310 \ \ 0.1792 & 1.670E+03 \ \ 1.041E+03 & K \\
\enddata
\end{deluxetable}



\end{document}